\documentclass[reprint,
superscriptaddress,amsmath,amssymb,aps, longbibliography
]{revtex4-2}

\usepackage[utf8]{inputenc}
\usepackage{multirow}

\usepackage[lining,semibold]{libertine} 
\usepackage[libertine, cmintegrals, bigdelims, vvarbb]{newtxmath}

\usepackage{amsmath}
\usepackage{amsfonts}
\usepackage{mathrsfs}
\usepackage{gensymb}
\usepackage{bbm}
\usepackage{dsfont}

\usepackage{chemformula}
\usepackage[caption=false]{subfig}
\usepackage{chemfig}
\usepackage[version=3]{mhchem}

\usepackage{tikz}
\usetikzlibrary{matrix,positioning,decorations.pathreplacing}

\usepackage{braket}

\usepackage{soul}
\usepackage{xcolor}

\usepackage{scalerel} 
\usepackage{comment}

\usepackage{blkarray}

\usetikzlibrary{arrows, automata}
\usepackage{algorithm}
\usepackage{algpseudocode}

\definecolor{webgreen}{rgb}{0,.5,0}
\definecolor{webbrown}{rgb}{.6,0,0}
\definecolor{grigio}{rgb}{.85,.85,.85} 
\definecolor{RoyalBlue}{rgb}{0.0, 0.14, 0.4}
\definecolor{skyblue1}{rgb}{0.45,0.62,0.81}
\definecolor{skyblue2}{rgb}{0.2,0.39,0.64}
\definecolor{skyblue3}{rgb}{0.13,0.29,0.53}
\definecolor{scarlet1}{rgb}{0.93,0.16,0.16}
\definecolor{scarlet2}{rgb}{0.8,0,0}
\definecolor{scarlet3}{rgb}{0.64,0,0}

\definecolor{g}{gray}{0.50}

\usepackage[normalem]{ulem}

\usepackage{hyperref}
\hypersetup{%
    colorlinks=true, linktocpage=true, pdfstartpage=1, pdfstartview=FitV,%
    breaklinks=true, pdfpagemode=UseNone, pageanchor=true, pdfpagemode=UseOutlines,%
    plainpages=false, bookmarksnumbered, bookmarksopen=true, bookmarksopenlevel=1,%
    hypertexnames=true, pdfhighlight=/O,
    urlcolor=webbrown, linkcolor=RoyalBlue, citecolor=webgreen, 
    pdftitle={},%
    pdfauthor={Benedikt Remlein},%
    pdfsubject={},%
    pdfkeywords={},%
    pdfcreator={pdfLaTeX},%
    pdfproducer={LaTeX REVTeX}%
}

\usepackage[capitalise]{cleveref}

\usepackage{verbatim}

\usepackage{kbordermatrix}

\usetikzlibrary{calc}
\newcommand{\crn}[2]{\ch{<=>[$#2$][$#1$]}}

\newcommand{\figwidth}{0.90\linewidth}

\usepackage{graphicx}
\usepackage{mathtools}
\usepackage{comment}

\begin{document}

\title{Singular Behavior of Observables at Hopf Bifurcations}

\author{Benedikt Remlein}
\email{benedikt.remlein@uni.lu}
\affiliation{Complex Systems and Statistical Mechanics, Department of Physics and Materials Science, University of Luxembourg, 30 Avenue des Hauts-Fourneaux, L-4362 Esch-sur-Alzette, Luxembourg}

\author{Massimiliano Esposito}
\email{massimiliano.esposito@uni.lu}
\affiliation{Complex Systems and Statistical Mechanics, Department of Physics and Materials Science, University of Luxembourg, 30 Avenue des Hauts-Fourneaux, L-4362 Esch-sur-Alzette, Luxembourg}

\date{\today}


\begin{abstract}
Hopf bifurcations are a universal route to self-sustained oscillations in driven systems. Despite the absence of any singular stationary state, we show that time-averaged observables generically exhibit singularities at the onset of oscillations.  
The origin of this behavior is geometric: phase averaging over the emergent periodic attractor eliminates odd powers of the oscillation amplitude, while the squared amplitude varies smoothly with the distance from the bifurcation. Consequently, the excess of a time-averaged observable admits an integer-power expansion; observables remain finite but display discontinuities in finite-order derivatives. This yields an Ehrenfest-like hierarchy of Hopf singularities, in which the first nonanalytic derivative is determined by the lowest-order coupling between the observable and the limit-cycle waveform that survives phase averaging. The time averages of generic observables therefore exhibit kink singularities, while symmetry or geometric cancellations can suppress lower-order couplings and shift nonanalyticity to higher derivatives.
We demonstrate this mechanism in chemical, electronic, and climate oscillators. Our results identify supercritical Hopf bifurcations as a universal mechanism for nonanalytic observable behavior, where singular yet non-diverging features emerge without any underlying singular stationary state. 
\end{abstract}

\maketitle

 
\textit{Introduction.—}
Nonanalytic behavior of measurable observables is a defining signature of critical phenomena \cite{gold92,landau}. 
In statistical systems, such singularities arise from qualitative changes of the relevant steady-state measure: in equilibrium through nonanalytic Gibbs ensembles, and out of equilibrium through critical stationary states or fluctuation-induced transitions in driven or absorbing systems \cite{marr99,hinr00}, as well as in broader contexts including complex and living systems \cite{mora11,muno18}. For many interacting driven systems, sharp transitions emerge most naturally in macroscopic limits in which fluctuations become subleading and the stationary measure concentrates on deterministic attractors \cite{kurt70,vankampen}. In this sense, bifurcations of mean-field dynamics provide a natural route to nonequilibrium mean-field criticality, a perspective developed in the theory of dissipative structures, where dynamical instabilities are interpreted as nonequilibrium phase transitions \cite{glansdorff,nicolis,fala25}.

Among dynamical instabilities, the Hopf bifurcation is one of the most ubiquitous routes to self-sustained oscillations in nonlinear science \cite{guck83,wigg03}. Such oscillatory instabilities underlie optical and electronic oscillators \cite{hake75,erne10,vand26,stro00}, fluid and chemical instabilities \cite{cros93,epst96}, biological rhythms including biochemical clocks, hair-cell dynamics, and neural activity \cite{nova08,gold10,cama00,mart01,fitz61,izhi06}, collective synchronization phenomena in coupled oscillators \cite{kuramoto,aceb05}, and climate variability \cite{jin97a,jin97b,wang18}. The onset of oscillations via a Hopf bifurcation constitutes a paradigmatic dynamical transition in nonequilibrium systems. However, a general theoretical framework connecting this transition to the behavior of measurable observables remains lacking.

Recent thermodynamically consistent oscillator models have shown that the onset of self-sustained oscillations can generate sharp thermodynamic signatures, including kinks in dissipation, work rates, and nonequilibrium free energies \cite{herp18,fala18,nguy18,meib24a,meib24,gopa25,ptas25c,chud26}. Related work has examined how coupling and synchronization reshape entropy production within already established oscillatory phases \cite{impa15,izum16,lee18,zhan20,burn26}. 
Recent studies have furthermore identified singular fluctuation and response behavior near nonequilibrium bifurcations \cite{iwat10,nguy20,reml24,ptas25a,ptas25b} including Hopf bifurcations in chemical reaction networks \cite{reml24thesis,tama26}.
These results raise a broader question: are singularities observed at the onset of oscillations model-specific thermodynamic features, or are they universal consequences of supercritical Hopf bifurcations for time-averaged observables?

We show that the latter is the case. For generic, non-degenerate supercritical Hopf bifurcations in deterministic finite-dimensional systems,
the singular behavior of smooth observables previously observed in specific models across different areas of physics is not model-specific, but rather reflects a generic observable-level consequence of such bifurcations.

For stationary bifurcations, observable singularities arise in a direct way: observables generically inherit their leading nonanalyticity from a critical stationary branch or from discontinuous branch selection.
A supercritical Hopf bifurcation is fundamentally different. The stationary branch remains smooth in the vicinity of the bifurcation, while a stable periodic orbit emerges, breaks continuous time-translation symmetry to periodic motion, and exhibits the characteristic square-root growth of its amplitude near the bifurcation. No singular stationary branch, in the sense of nonanalytic parameter dependence or discontinuous branch selection, is therefore available from which observable nonanalyticities could be inherited.
Thus the standard mechanism for generating observable singularities is absent. Nevertheless, we show that time-averaged observables are generically nonanalytic at the onset of oscillations despite the absence of any singular stationary state.

The mechanism is not inheritance from a singular stationary state, but phase averaging over the emergent oscillatory attractor. 
After radial relaxation, the attractor has fixed amplitude and a single phase degree of freedom, so time averages reduce to phase averages. This averaging removes all odd orders in the oscillatory displacement from the smoothly continued stationary branch. 
Consequently, the excess of a time-averaged observable over its smoothly continued stationary value admits an integer-power expansion near the bifurcation, while remaining zero below it.

The first nonvanishing term determines the leading singular derivative. Generically this produces a kink, whereas symmetry or geometric cancellations shift the singularity to higher derivative order. This yields an Ehrenfest-like hierarchy of Hopf singularities classified by the first nonanalytic derivative. Hopf criticality is therefore singular without divergence: observables remain continuous and finite, while finite-order derivatives become discontinuous. This places previously observed thermodynamic kinks near oscillatory transitions into a broader universal framework.

These results identify a distinct observable-level mechanism of dynamical criticality. Stationary bifurcations transmit singularities through critical stationary states, whereas Hopf bifurcations generate them through phase averaging over an emergent oscillatory attractor. We demonstrate the theory for thermodynamically consistent chemical and electronic oscillators, and for a canonical reduced climate oscillator model, showing that Hopf bifurcations constitute a distinct class of singular but non-divergent nonequilibrium critical phenomena.

\begin{figure}[t]
\centering
\includegraphics[width=\linewidth]{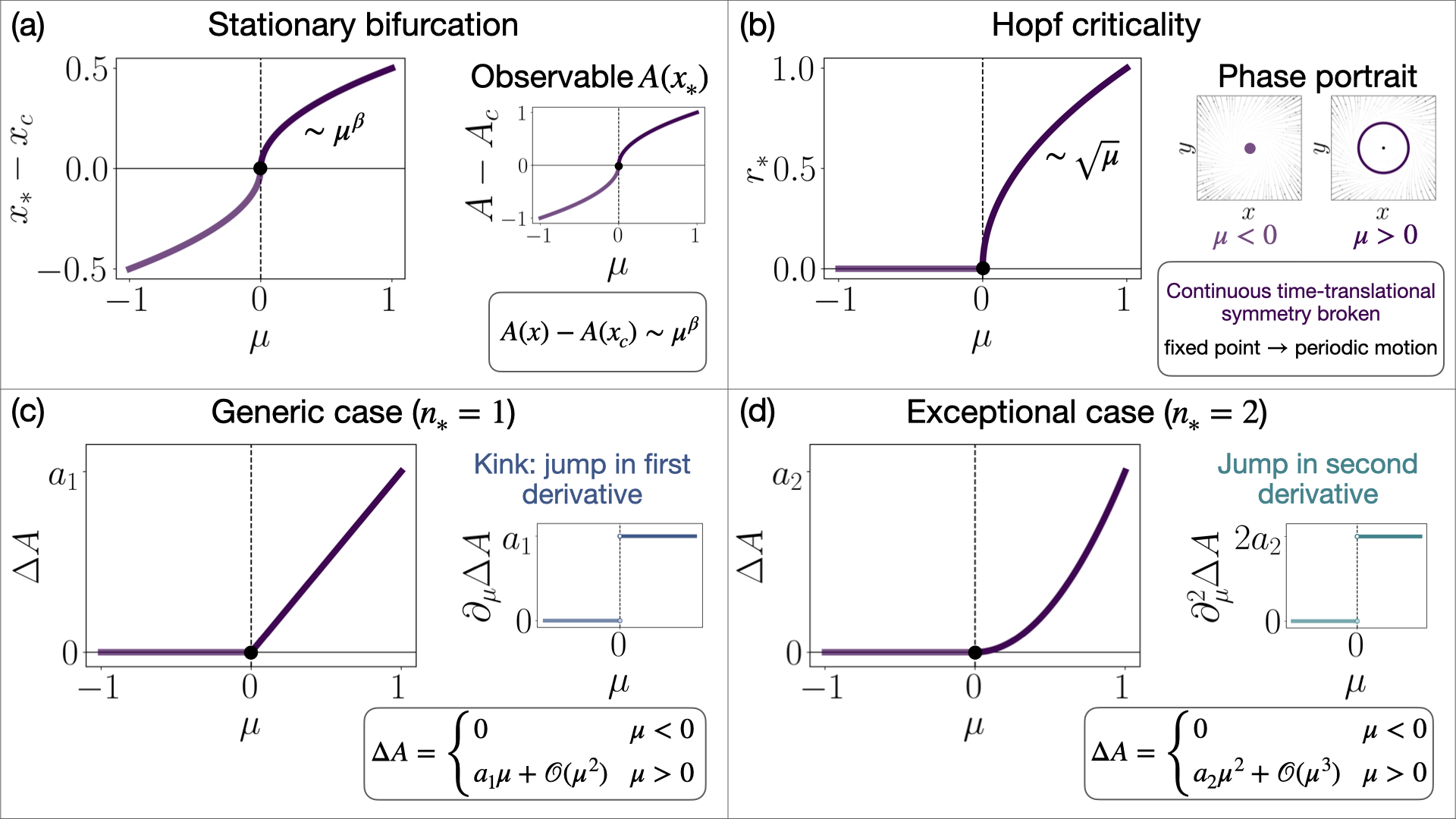}
\caption{
Observable singularities at bifurcations.
(a) Stationary bifurcations transmit nonanalytic behavior through the critical stationary branch \(x_\ast(\mu)\). Observables inherit the same leading singularity under composition, e.g.,
\(
A(x_\ast(\mu))-A(x_c)\sim \mu^\beta
\)
for
\(
x_\ast(\mu)-x_c\sim \mu^\beta
\),
or inherit jumps under discontinuous branch selection (not shown).
(b) Supercritical Hopf bifurcation. The stationary branch remains smooth, while a stable limit cycle emerges for \(\mu>0\) with amplitude
\(
r_\ast\sim \mu^{1/2}
\).
The phase portraits illustrate the transition from a stable fixed point (\(\mu<0\)) to periodic motion (\(\mu>0\)), corresponding to spontaneous breaking of continuous time-translation symmetry.
(c) Generic Hopf case for the excess observable
\(
\Delta A=\langle A\rangle-A_\ast(\mu)
\),
where
\(
A_\ast(\mu)=A(x_\ast(\mu))
\).
One finds
\(
\Delta A=0
\)
for \(\mu<0\) and
\(
\Delta A=a_1\mu+\mathcal O(\mu^2)
\)
for \(\mu>0\), producing a kink. The inset shows the jump in the first derivative.
(d) Cancellation case \(a_1=0\). The leading term is shifted to
\(
\Delta A=a_2\mu^2+\mathcal O(\mu^3)
\),
so the observable itself and its slope remain continuous, while the second derivative is discontinuous (inset).
}
\label{fig:overview}
\end{figure}

\textit{Observable inheritance at stationary bifurcations.—}
Let \(x_\ast(\mu)\in\mathbb R^n\), \(n\ge1\), denote a stable stationary branch of a dynamical system controlled by the parameter \(\mu\),
and let \(A(x)\) be a smooth scalar observable for which the local expansions used throughout are well defined.
For stationary bifurcations, observables inherit singularities directly from the critical branch: if the transition is continuous,
\begin{equation}
x_\ast(\mu)-x_c\sim \mu^\beta\,,
\end{equation}
then generically
\begin{equation}
A(x_\ast(\mu))-A(x_c)\sim \mu^\beta\,,
\end{equation}
by Taylor expansion, where \(x_c \equiv x_\ast(0)\) denotes the critical value, see Fig.~\ref{fig:overview}a.
If stability switches discontinuously between coexisting branches, \(A(x)\) inherits the corresponding jump. 


\textit{Observable singularities at Hopf bifurcations.—}
Consider a smooth finite-dimensional dynamical system
\begin{equation}
\dot x=f(x,\mu)\,,
\label{eq:dynamicalSystem}
\end{equation}
undergoing a generic nondegenerate supercritical Hopf bifurcation at \(\mu=0\). For \(\mu<0\), a stable fixed point \(x_\ast(\mu)\) exists; for \(\mu>0\), a stable limit cycle emerges continuously from it, while the fixed-point branch remains smooth in the vicinity of the bifurcation.

Let \(A(x)\) be a sufficiently smooth scalar observable and define
\begin{equation}
A_\ast(\mu) \equiv A(x_\ast(\mu))\,,
\end{equation}
together with the excess observable
\begin{equation}
\Delta A(\mu)\equiv \langle A\rangle-A_\ast(\mu)\,,
\label{eq:defDelta}
\end{equation}
where \(\langle\cdot\rangle\) denotes averaging over one period of the stable limit cycle for \(\mu>0\).

Then, sufficiently close to the bifurcation,
\begin{equation}
\Delta A(\mu)=0\,,
\qquad \mu<0\,,
\end{equation}
while for \(\mu>0\),
\begin{equation}
\Delta A(\mu)=\sum_{m\ge1} a_m \mu^m\,.
\label{eq:main}
\end{equation}
Hence observables remain finite at a Hopf bifurcation, and criticality appears through a finite-order derivative singularity rather than a divergence.

The first nonzero coefficient determines the singular response. Defining
\begin{equation}
n_\ast=\min\{m\ge1:\,a_m\neq0\}\,,
\label{eq:defnstar}
\end{equation}
the first \(n_\ast-1\) derivatives of \(\Delta A\) remain continuous at \(\mu=0\), whereas the \(n_\ast\)th derivative is discontinuous. Geometrically, \(n_\ast\) identifies the lowest-order coupling between the observable and limit-cycle waveform that survives phase averaging. The generic case \(n_\ast=1\) corresponds to sensitivity already at order \(r_\ast^2\sim\mu\), while larger \(n_\ast\) indicates that lower-order couplings are suppressed by symmetry or geometric cancellations.

Thus generic supercritical Hopf bifurcations generate an Ehrenfest-like hierarchy of observable singularities classified by the first nonanalytic derivative. In particular, for \(n_\ast = 1\) the leading contribution is linear in the control parameter, \(\Delta A = a_1 \mu + \dots\), while for \(n_\ast = 2\) the observable exhibits a quadratic onset, see Fig.~\ref{fig:overview}.


\textit{Origin of the integer-power law.—}
By center-manifold reduction, the dynamics near the bifurcation are locally two-dimensional \cite{guck83,craw91,wigg03}. In suitable amplitude-phase \((r,\theta)\) coordinates, the dynamics is described by
\begin{equation}
\dot r=rF(r^2,\mu)\,,
\qquad
\dot\theta=G(r^2,\mu)\,,\label{eq:vectorFieldRTheta}
\end{equation}
with \(F\) and \(G\) smooth and no explicit phase dependence, see Sec.~1 of the Supplemental Material (SM) \cite{SM} for a justification of Eq. (\ref{eq:vectorFieldRTheta}).

For \(\mu>0\), the attracting cycle has constant amplitude
\begin{equation}
r=r_\ast(\mu)\,,
\end{equation}
with
\begin{equation}
r_\ast^2(\mu)=c_1\mu+c_2\mu^2+\ldots\,,
\qquad c_1>0\,.
\label{eq:rSquaredExpansion}
\end{equation}

The oscillating trajectory can be written as
\begin{equation}
x(\theta,\mu)=x_\ast(\mu)+\delta x(\theta,\mu)\,,
\end{equation}
where \(\delta x\) is a smooth, \(2\pi\)-periodic function of \(\theta\), satisfies
\(
\delta x=\mathcal O(r_\ast)
\),
and describes the local waveform of the emergent oscillation in state space.

Since \(A\) is locally expandable around \(x_\ast\), the difference \(A(x_\ast+\delta x)-A(x_\ast)\) admits a Taylor expansion in the oscillatory displacement \( \delta x\). Because \(\delta x\) is periodic, it possesses an amplitude-resolved Fourier expansion, which transfers to the expansion of \(A(x_\ast+\delta x)-A(x_\ast)\). Averaging over one period removes all nonzero Fourier modes and, as shown in Appendix~\ref{app:odd}, eliminates all odd orders in \(r_\ast\).

Hence only even powers survive:
\begin{equation}
\Delta A(\mu)=\sum_{j\ge1} b_j(\mu)\,r_\ast(\mu)^{2j}\,,
\end{equation}
with smooth coefficients \(b_j(\mu)\). Using Eq.~(\ref{eq:rSquaredExpansion}) and the expansion of \(b_j(\mu)\) in \(\mu\) then yields the ordinary one-sided power series in Eq.~(\ref{eq:main}). The coefficients \(b_j(\mu)\) collect the symmetry-allowed couplings between the local derivatives of the observable and the Fourier components of the emergent waveform. The Ehrenfest hierarchy is therefore determined by the lowest-order coupling that remains nonzero after phase averaging.

The coefficients \(a_m\) are constructive: they are determined by the Hopf normal form, the center-manifold embedding, and local derivatives of the observable.
In particular, the leading-order coefficient is given by
\begin{equation}
a_1
=
-\frac{\kappa}{l_1}
\left\langle
\nabla A(x_c)\cdot X_2(\theta)
+
\frac{1}{2}
X_1(\theta)^{\mathsf T}
H_A(x_c)
X_1(\theta)
\right\rangle \,,
\label{eq:main_text_a1}
\end{equation}
where \(H_A(x_c)\) is the Hessian of \(A\) evaluated at the bifurcation, \(X_1\) and \(X_2\) denote the first- and second-order contributions to the local limit-cycle waveform \(\delta x\), \(\kappa\) is the rate at which the real part of the critical eigenvalue crosses zero as \(\mu\) is varied, and \(l_1\) is the first Lyapunov coefficient. For the supercritical Hopf bifurcation considered here, \(-\kappa/l_1>0\). The coefficient therefore factorizes into a dynamical amplitude contribution fixed by the local Hopf normal form and a geometric contribution describing how the observable couples to the emerging waveform. The definitions of \(X_1\) and \(X_2\), together with the derivation of Eq.~\eqref{eq:main_text_a1}, are given in Appendix~\ref{app:practical_hierarchy}.

\begin{figure*}[t!]
\centering
\includegraphics[width=\linewidth]{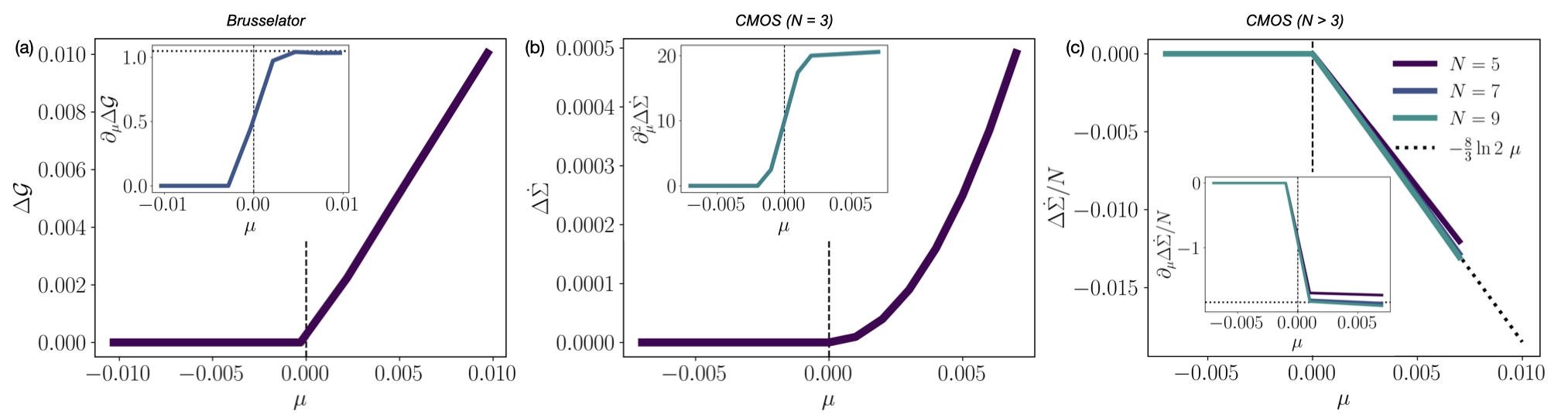}
\caption{
Observable singularities in the vicinity of a Hopf bifurcation.
Main panels show excess observables, while insets display the lowest derivative that becomes discontinuous at the bifurcation.
(a) Reversible Brusselator: the excess semi-grand Gibbs free energy belongs to the generic kink class. The inset shows the first derivative; the dotted line indicates the local Hopf prediction \(g_1=1.05\). The system parameters are (in arbitrary units)
\(k_1=0.1\), \(k_{-1}=1\), \(k_2=1\), \(k_{-2}=1\), \(k_3=0.1\), \([A]=1\), and \([B]=3\), while \(k_{-3}\) is determined by Eq. (\ref{eq:LDB}) as function of the control parameter \(\mathcal A\).
(b) CMOS ring oscillator (\(N=3\)): cancellation of the linear term shifts the singularity to the second derivative.
(c) CMOS ring oscillator for odd \(N>3\): the excess entropy production per stage, shown for \(N=5,7,9\), exhibits the generic kink class. The dotted line indicates the large-\(N\) asymptotic slope \(-(8/3)\ln2\).
Vertical dashed lines mark the Hopf bifurcation at \(\mu=0\).
}
\label{fig:examples}
\end{figure*}

\textit{Examples.—}
We illustrate the theory for three representative classes of oscillatory systems spanning chemistry, electronics, and climate dynamics.
For each case, we solve the corresponding dynamical system numerically and examine the excess observable \(\Delta A(\mu)\) while varying the control parameter \(\mu\), Fig.~\ref{fig:examples}.

\textit{Chemical oscillator: reversible Brusselator.—}
We consider the reversible Brusselator reaction network \cite{schn79,qian02,nguy18,reml22}
\begin{equation}
A \crn{-1}{1} X\,,\qquad
3X \crn{2}{-2} 2X+Y\,,\qquad
Y \crn{3}{-3} B\,,
\label{eq:CRNBruss}
\end{equation}
where the interconversion of the internal species \(X\) and \(Y\) is driven by chemostatted species \(A\) and \(B\), whose concentrations \([A]\) and \([B]\) are held fixed. The detailed analysis is done in Sec. 4 of SM \cite{SM} and we now report its main findings.
Under mass-action kinetics \cite{gasp04}, the concentration dynamics undergo a supercritical Hopf bifurcation as the chemical affinity driving the system out-of-equilibrium is varied
\begin{equation}
\mathcal A\equiv\ln\frac{[B]k_3k_2k_{-1}}{[A]k_1k_{-2}k_{-3}}\,, \label{eq:LDB}
\end{equation}
where \(k_\rho\) denotes the kinetic constant of reaction \(\rho\in\{\pm1,\pm2,\pm3\}\). 
This parametrization ensures a thermodynamically consistent description \cite{schm06,rao16,rao18}, leading to the Kelvin-like second law \(\dot \Sigma = \dot W - d_t \mathcal G \geq 0\), 
where \(\dot \Sigma\) is the nonnegative entropy production rate, $\dot W$ the nonconservative work rate driving the system out of equilibrium, and \(\mathcal G\) the semi-grand Gibbs free energy.
Here \(\dot W=-\mathcal A\,J_1\), where \(J_1 \equiv k_1 [A] - k_{-1} [X]\) measures the exchange current with the reservoir of species $A$ and \([X]\) is the concentration of species \(X\).
In a stationary or periodic state, since \(\langle d_t \mathcal G \rangle = 0 \), we find that $\langle \dot \Sigma\rangle =  \langle \dot W \rangle$.
Using the distance to the bifurcation \(\mu=\mathcal A-\mathcal A_c\),
numerical time averaging shows a clear kink in \(\Delta\mathcal G\), placing the observable in the generic class
\(n_\ast^{\mathcal G}=1\),
see Fig.~\ref{fig:examples}a. Moreover, numerically evaluating Eq. (\ref{eq:main_text_a1}) yields
\begin{equation}
\Delta\mathcal G=1.05\,\mu+\mathcal O(\mu^2)\,.
\end{equation}
The direct trajectory average and the local coefficient evaluation are in good agreement, confirming the predicted kink amplitude, see inset of Fig. \ref{fig:examples}a.
The time-averaged entropy production rate, likewise exhibits the generic kink singularity,  \(n_\ast^{\dot \Sigma}=1\), in contrast to the geometry-induced cancellation discussed below for the three-stage CMOS oscillator.

\textit{Electronic oscillator: CMOS ring oscillator.—}
We next consider an \(N\)-stage CMOS ring oscillator, a driven dissipative electronic system in which nonequilibrium currents across coupled inverters generate self-sustained voltage oscillations \cite{frei21b,gopa25}. The detailed analysis is done in Sec. 5 of SM \cite{SM}.
The dimensionless output voltages \(x_i\) obey the dynamics 
\begin{equation}
\dot x_i=
2\left[
\sinh(x_{i-1}-x_i)-e^{a}\sinh(x_{i-1})
\right]\,,
\qquad i=1,\dots,N\,,\label{eq:CMOSvec}
\end{equation}
with cyclic indexing, where \(a\) quantifies the nonequilibrium electrical drive relative to thermal fluctuations. Following Ref.~\cite{gopa25}, we use the rescaled distance from the Hopf bifurcation
\(\mu=a-a_c\) as control parameter with \(a_c \equiv \log[1  + \sec \pi/N]\). As \(\mu\) crosses zero, the stationary operating point loses stability through a supercritical Hopf bifurcation, giving rise to self-sustained oscillations.
Those oscillations are maintained by a constant nonequilibrium drive across each CMOS inverter, leading to the entropy production rate
\begin{equation}
    \dot \Sigma \equiv 2a \sum_i [e^a \cosh(x_{i-1}) - \cosh(x_i - x_{i-1})]\,.\label{eq:EPRCMOS}
\end{equation}

Due to the anti-symmetry of the vector-field (\ref{eq:CMOSvec}), the quadratic mean-shift contribution of the Hopf waveform vanishes, \(\left \langle  X_{2}(\theta)\right\rangle=0\). Consequently, the observable-geometric contribution to the leading coefficient in Eq.~(\ref{eq:main_text_a1}) is determined solely by the Hessian of the entropy production rate restricted to the critical Hopf eigenspace, as discussed in Appendix \ref{app:practical_hierarchy}.
Numerically evaluating Eq. (\ref{eq:EPRCMOS}) for the three-stage circuit (\(N=3\)), we find that \(n_\ast^{\dot\Sigma}=2\), see Fig.~\ref{fig:examples}b.
Ref.~\cite{gopa25} already identified the cancellation of the linear contribution for \({N=3}\); here we show that this cancellation implies a quadratic Hopf singularity. In the language of the hierarchy introduced above, the leading-order coupling between the observable and the limit-cycle waveform is suppressed, so that the first surviving singular contribution appears only at second order.

Thus, for the particular observable geometry realized by the three-stage circuit (\( N = 3\)), the entropy production rate exhibits no quadratic curvature along the critical Hopf mode.

For odd \(N>3\), the small-\(\mu\) expansion of the entropy production rate, Eq.~(43) of Ref.~\cite{gopa25}, yields
\begin{equation}
s_1^{(N)}
=
-4N\ln\!\left(1+\sec\frac{\pi}{N}\right)
\left[
1-\frac{1}{
\cos(\pi/N)\left(1+2\cos(\pi/N)\right)}
\right]\,,
\end{equation}
with \(\Delta\dot\Sigma=s_1^{(N)}\mu+\mathcal O(\mu^2)\), thus placing larger odd rings in the generic kink class \(n_\ast^{\dot\Sigma}=1\).
Fig.~\ref{fig:examples}c shows the excess entropy production per stage for \(N=5,7,9\), together with the corresponding jump in the first derivative (inset), directly visualizing the kink singularity.
In the large-\(N\) limit,
\begin{equation}
\frac{s_1^{(N)}}{N}
=
-\frac{8}{3}\ln 2
+
\frac{2\pi^2}{9N^2}(5\ln2-3)
+
\mathcal O(N^{-4})\,,
\end{equation}
so the kink strength grows extensively with circuit size, while the slope per stage approaches the asymptote \(-(8/3)\ln2\), shown by the dotted line in Fig.~\ref{fig:examples}c. The convergence of \(s_1^{(N)}/N\) demonstrates that the kink persists as an intensive nonanalyticity in the many-body limit, rather than being a finite-size effect. At the same time, the three-stage circuit (\(N=3\)) exhibits a quadratic onset due to cancellation of the linear term. Thus, even for a fixed observable, here the entropy production rate, and bifurcation type, both the realized Hopf singularity class and its amplitude can vary within a single oscillator family.

\textit{Climate oscillator: ENSO recharge--discharge model.—}
As a geophysical example, we consider the canonical recharge--discharge oscillator description of the El Niño--Southern Oscillation (ENSO) \cite{jin97a,jin97b,wang18},
\begin{align}
\dot T &= C T + D h - \varepsilon T^3\,,\\
\dot h &= -E T - R_h h\,,
\end{align}
where \(T\) denotes the eastern-Pacific sea-surface-temperature anomaly and \(h\) the western-Pacific thermocline-depth anomaly. In the recharge--discharge mechanism, subsurface heat-content adjustments provide a principal negative feedback terminating ENSO growth and sustaining interannual oscillations \cite{wang18}. The climatological stationary state loses stability through a supercritical Hopf bifurcation at
\(C_c=R_h\) and \(\omega_c^2=DE-R_h^2>0\), where \(C\) quantifies the ocean--atmosphere feedback that amplifies temperature anomalies and therefore acts as the natural control parameter of the instability.
Taking \(\mu=C-C_c\) as control parameter leads to
\begin{equation}
\Delta T^2=
\frac{2}{3\varepsilon}\mu+\mathcal O(\mu^2)\,.
\end{equation}
The derivation is presented in Sec. 6 of the SM \cite{SM}.
Thus the ENSO recharge--discharge oscillator belongs to the generic kink class with singularity order
\(
n_\ast^{T^2}=1
\).


\textit{Conclusion.—}
We have shown that supercritical Hopf bifurcations constitute a distinct class of non-divergent critical behavior and, in thermodynamically consistent driven systems, provide a natural realization of nonequilibrium mean-field phase transitions. Although the stationary branch remains smooth and no singular stationary state exists, sufficiently smooth time-averaged observables are nevertheless generically nonanalytic at the onset of oscillations. These singularities do not arise from stationary criticality, but from phase averaging over the emergent periodic attractor.

This mechanism converts the amplitude expansion of the emerging limit cycle—whose leading behavior exhibits square-root scaling—into an integer-power expansion in the distance from the bifurcation. As a result, time-averaged observables remain finite at criticality, while finite-order derivatives become discontinuous. Generically, the leading signature is a kink, whose amplitude can be computed explicitly from local Hopf data, whereas symmetry or geometric cancellations can suppress lower-order observable--waveform couplings and shift the singularity to higher derivative order, yielding an Ehrenfest-like hierarchy. The singular coefficients therefore reflect how different observables probe the local geometry of the emerging oscillation, including the quadratic mean deformation of the cycle and the structure of the critical eigenspace.

For thermodynamically consistent driven systems, this provides a general explanation for sharp features previously observed in free-energy-like functions, dissipation, and work rates near oscillatory transitions. Our chemical and electronic examples show that such signatures are not model-specific, but follow from the local Hopf structure together with observable smoothness and time averaging.

The framework is not restricted to thermodynamic observables; it also applies to moments and quantities constructed smoothly from them, including variances, oscillation powers, and mean-square amplitudes. The ENSO recharge--discharge model illustrates this broader scope by realizing the same kink singularity in the variance of the sea-surface-temperature anomaly.

The CMOS ring oscillator further demonstrates that the singularity class is not fixed by the observable or by the Hopf bifurcation alone. The same observable, for example the entropy production rate, can exhibit either generic kink behavior or a higher-order onset depending on the interplay between the symmetry structure of the underlying dynamics and the geometry of the observable. 
Hopf universality therefore does not imply a unique singularity class.

More generally, these results show that Hopf bifurcations can induce nonanalytic behavior in measurable observables without singular stationary order parameters. 
The present framework applies to observables that admit a local expansion around the fixed point and therefore does not include non-smooth quantities such as the oscillation amplitude itself, which instead exhibits the familiar square-root scaling in the vicinity of a supercritical Hopf bifurcation.

Extending this framework to noisy finite-size systems and spatially extended media is a natural direction for future work.


\begin{acknowledgments}
This research is supported by Project
INTER/ANR/25/19593353-NERD, funded by Fonds National de la Recherche (FNR) Luxembourg and Agence Nationale de la Recherche (ANR) France, and by Project NEQPHASETRANS (C24/MS/18933049) funded by FNR.
\end{acknowledgments}


\begin{center}
\large\bf End Matter
\end{center}
\vspace{-0.8cm}

\appendix


\section{Phase averaging and cancellation of odd powers}
\label{app:odd}

We justify the key structural result underlying the main text, namely that phase averaging over the emergent limit cycle eliminates all odd powers of the Hopf amplitude and thereby produces an ordinary integer-power expansion of the excess observable in the distance from the bifurcation. We summarize only the essential structural steps; a detailed derivation including the explicit Fourier bookkeeping is provided in Sec.~1 and 2 of the SM \cite{SM}.

Near the Hopf bifurcation, the oscillatory trajectory admits a smooth expansion in the complex Hopf amplitude
\begin{equation}
z=r_\ast e^{i\theta}\,,
\qquad
\bar z=r_\ast e^{-i\theta}\,,
\end{equation}
through the center-manifold embedding 
\begin{equation}
x(\theta,\mu)-x_\ast(\mu)
=
\Phi(z,\bar z,\mu)\,,
\end{equation}
where \(\Phi\) is smooth in \(z,\bar z,\mu\). Expanding in powers of \(z\) and \(\bar z\) gives
\begin{equation}
\delta x(\theta,\mu)
=
\sum_{l,k\ge0}
\phi_{lk}(\mu)
e^{i(l-k)\theta}
r_\ast(\mu)^{\,l+k}\,,
\label{eq:app_dx_short}
\end{equation}
where \(\phi_{lk}(\mu)\) denote the expansion coefficients of \(\Phi(z,\bar z,\mu)\).

For an observable \(A(x)\), a Taylor expansion gives
\begin{equation}
A(x_\ast+\delta x)-A(x_\ast)
=
\sum_{m\ge1}
\frac1{m!}
D^mA(x_\ast)[\delta x,\ldots,\delta x]\,,\label{eq:app:taylor}
\end{equation}
where \(D^mA \) denotes the \(m\)-th order derivative tensor of observable \( A \). Substituting Eq.~(\ref{eq:app_dx_short}) into Eq.~(\ref{eq:app:taylor}) produces terms proportional to
\begin{equation}
e^{iq\theta}r_\ast(\mu)^p\,,
\end{equation}
where
\begin{equation}
q\equiv\sum_i(l_i-k_i)\,,
\qquad
p\equiv\sum_i(l_i+k_i)\,.
\end{equation}
Phase averaging removes all terms with \(q\neq0\). The surviving condition \(q=0\) implies
\(
\sum_i l_i=\sum_i k_i\,,
\)
and therefore
\(
p=2\sum_i l_i\,,
\)
which is necessarily even. Hence only even powers of \(r_\ast\) survive after averaging:
\begin{equation}
\Delta A(\mu)
=
\sum_{j\ge1}
b_j(\mu)r_\ast(\mu)^{2j}\,.
\end{equation}

Since \(r_\ast^2(\mu)\) admits a one-sided expansion in \(\mu\) starting at the linear order, Eq. (\ref{eq:rSquaredExpansion}), and \(b_j(\mu)\) is smooth, the excess observable possesses the integer-power expansion
\begin{equation}
\Delta A(\mu)
=
\sum_{m\ge1}a_m\mu^m\,.
\end{equation}


\section{Leading Hopf coefficient}
\label{app:practical_hierarchy}

We summarize the structure of the leading-order coefficient in the excess observable expansion
\begin{equation}
\Delta A(\mu)=a_1\mu+\mathcal O(\mu^2)\,,
\qquad \mu>0\,,
\end{equation}
for a generic supercritical Hopf bifurcation. A detailed derivation is given in Sec.~3 of the SM \cite{SM}.

Near the bifurcation, the center-manifold dynamics reduce to the Hopf normal form
\begin{equation}
\dot r=\kappa\mu r+l_1 r^3+\ldots\,,
\end{equation}
where \(\kappa=\alpha'(0)\) and \(l_1\) is the first Lyapunov coefficient \cite{guck83,craw91,wigg03}. The stable limit-cycle amplitude therefore satisfies
\begin{equation}
r_\ast^2=c_1\mu+\mathcal O(\mu^2)\,,
\qquad
c_1=-\frac{\kappa}{l_1}\,.
\label{eq:app_compact_c1}
\end{equation}

The emerging oscillation admits the expansion
\begin{equation}
x(\theta,\mu)-x_\ast(\mu)
=
r_\ast X_1(\theta)
+
r_\ast^2 X_2(\theta)
+
\mathcal O(r_\ast^3)\,,
\end{equation}
where
\begin{equation}
X_1(\theta)=X_{1,c}\cos\theta+X_{1,s}\sin\theta
\end{equation}
spans the critical Hopf eigenspace. 
The second-order correction has the harmonic structure
\begin{equation}
X_2(\theta)
=
X_{2,0}
+
X_{2,c}\cos 2\theta
+
X_{2,s}\sin 2\theta\,,
\end{equation}
where \(X_{2,0}\) denotes the phase-independent quadratic mean shift of the emerging cycle, which is determined by
\begin{equation}
X_{2,0}
=
-\frac14L(0)^{-1}
\left[
F_2(X_{1,c},X_{1,c})
+
F_2(X_{1,s},X_{1,s})
\right]\,,
\label{eq:app_compact_X20}
\end{equation}
where \(F_2\equiv D_x^2f(x_c,\mu=0)\) and \(x_c\equiv x_\ast(0)\).

Expanding an observable \(A(x)\) around the smoothly continued stationary branch \(x_\ast(\mu)\) and averaging over one period yields
\begin{equation}
\Delta A=b_1(0)\,r_\ast^2+\mathcal O(r_\ast^4)\,,
\end{equation}
with
\begin{equation}
b_1(0)=
\nabla A(x_c)\cdot X_{2,0}
+
\frac14
\left(
X_{1,c}^\top H_A(x_c) X_{1,c}
+
X_{1,s}^\top H_A(x_c) X_{1,s}
\right)\,.
\label{eq:app_compact_b1}
\end{equation}
Combining Eqs.~(\ref{eq:app_compact_c1}) and (\ref{eq:app_compact_b1}) gives
\begin{equation}\small
a_1=
\left[
\nabla A(x_c)\cdot X_{2,0}
+
\frac14
\left(
X_{1,c}^\top H_A(x_c) X_{1,c}
+
X_{1,s}^\top H_A(x_c) X_{1,s}
\right)
\right]
\left(-\frac{\kappa}{l_1}\right).
\label{eq:app_compact_a1}
\end{equation}

We make a few comments about the above derivation:

\textit{i)} From Eq.~(\ref{eq:app_compact_a1}), the leading coefficient splits into a mean-shift contribution \(\nabla A(x_c)\cdot X_{2,0}\) and a curvature contribution involving the Hessian \(H_A(x_c)\). In general, both terms contribute and the jump direction is not fixed by convexity alone.
If the mean-shift term vanishes, the leading coefficient is determined entirely by the quadratic form of the Hessian. In this case, locally convex (concave) observables produce upward (downward) kinks.
For anti-symmetric vector fields satisfying \(f(-x,\mu)=-f(x,\mu)\), such as the CMOS example, one has \(F_2=0\) and hence \(X_{2,0}=0\). The leading singularity is then governed purely by the local curvature of the observable. In contrast, for generic observables---including nonequilibrium thermodynamic potentials or Lyapunov functions---both contributions can enter, and no universal jump direction follows from convexity alone.

\textit{ii)} The leading singular coefficient \(a_1\) is already fixed by the back-transformation of the leading-order Hopf cycle
\begin{equation}
z(\theta)\simeq \sqrt{c_1\mu}\,e^{i\theta}\,,
\end{equation}
into the original variables. The resulting waveform components \(X_1(\theta)\) and \(X_2(\theta)\) completely determine the leading coefficient, while higher harmonics, waveform distortions, and higher-order center-manifold corrections contribute only at order \(\mathcal O(\mu^2)\) and beyond.


\bibliography{refs}


\clearpage

\setcounter{section}{0}
\setcounter{subsection}{0}
\setcounter{equation}{0}
\setcounter{figure}{0}
\setcounter{table}{0}
\setcounter{page}{1}

\makeatletter
\renewcommand\thesection{S\arabic{section}}
\renewcommand\thesubsection{\thesection.\arabic{subsection}}

\renewcommand\theequation{\thesection.\arabic{equation}}
\numberwithin{equation}{section}

\renewcommand\thefigure{S\arabic{figure}}
\renewcommand\thetable{S\arabic{table}}

\def\appendixname{}
\def\sectionname{}
\makeatother

\begin{center}
\large\textbf{Supplemental Material}
\end{center}

\vspace{0.5cm}




\section{Local reduction near supercritical Hopf bifurcation}
\label{supp:hopf}

We summarize the standard reduction underlying the Hopf analysis in the main text.
Consider a smooth finite-dimensional dynamical system
\begin{equation}
\dot x=f(x,\mu)\,,
\qquad x\in\mathbb R^n\,,
\end{equation}
with a stationary branch \(x_\ast(\mu)\) undergoing a generic nondegenerate supercritical Hopf bifurcation at \(\mu=0\).

In the vicinity of the bifurcation, the Jacobian
\begin{equation}
L(\mu)=D_x f(x_\ast(\mu),\mu)\label{eq:app:Jacobian}
\end{equation}
has one simple complex-conjugate eigenvalue pair crossing the imaginary axis, \cite{guck83,craw91,wigg03}
\begin{equation}
\lambda_\pm(\mu)=\alpha(\mu)\pm i\omega(\mu)\,,\label{eq:app:eigenvalues}
\end{equation}
with
\begin{equation}
\alpha(0)=0\,,\qquad
\omega(0)=\omega_c\neq0\,,\qquad
\alpha'(0)\neq0\,.
\end{equation}
All remaining eigenvalues retain negative real part.

By the center-manifold theorem, the local dynamics reduce smoothly to a two-dimensional invariant manifold parameterized by a complex amplitude \(z\in\mathbb C\) \cite{guck83,craw91,wigg03}.

Following the procedure of the mapping to the normal form of a Hopf bifurcation, the reduced vector field can be written in the \(S^1\)-invariant form \cite{wigg03,craw91}
\begin{equation}
\dot z = z\,H(|z|^2,\mu)\,,
\label{eq:app_structural_nf2}
\end{equation}
where \(H\) is a smooth complex-valued function.
Under phase shifts \(z\mapsto e^{i\phi}z\), the reduced dynamics remain invariant, so the vector field depends on phase only through the overall factor \(z\).

Expanding Eq.~(\ref{eq:app_structural_nf2}) for small amplitude yields the familiar cubic truncation
\begin{equation}
\dot z=
(\alpha(\mu)+i\omega(\mu))z
+
\ell_1(\mu) z|z|^2
+
\mathcal O(5)\,,
\end{equation}
where \(\ell_1(\mu)\in\mathbb C\) is a generally complex coefficient and \(\mathcal O(5)\) denotes terms of total order at least five in \((z,\bar z)\). 

Writing
\begin{equation}
z=r e^{i\theta}\,,
\end{equation}
Eq.~(\ref{eq:app_structural_nf2}) becomes
\begin{equation}
\dot r = r\,F(r^2,\mu)\,,
\qquad
\dot\theta = G(r^2,\mu)\,,
\label{eq:app_polar_nf2}
\end{equation}
with no explicit \(\theta\) dependence, where \(F\) and \(G\) are smooth real-valued functions.
Thus the amplitude evolves autonomously, while the phase drifts around the cycle.

For a supercritical Hopf bifurcation, the stable periodic orbit corresponds to a nonzero root of
\begin{equation}
F(r_\ast^2,\mu)=0\,.
\end{equation}
Defining
\begin{equation}
s=r_\ast^2\,,
\end{equation}
this becomes
\begin{equation}
F(s,\mu)=0\,.
\end{equation}

For a generic nondegenerate Hopf bifurcation, the first Lyapunov coefficient, \(l_1 \equiv \operatorname{Re}\ell_1(0)\), is nonzero, i.e.,
\begin{equation}
\partial_sF(0,0) = l_1\neq0\,.
\end{equation}
Hence, by the implicit-function theorem, there exists a smooth one-sided branch \(s=s(\mu)\),
for \(\mu>0\), satisfying
\begin{equation}
s(\mu)=c_1\mu+c_2\mu^2+c_3\mu^3+\ldots\,,
\qquad c_1>0\,.
\label{eq:app_smooth_s}
\end{equation}
This leads to
\begin{equation}
r_\ast(\mu)=\sqrt{s(\mu)}
=
\sqrt{c_1}\,\mu^{1/2}+\mathcal O(\mu^{3/2})\,,
\end{equation}
thus, the usual amplitude scaling \(r_\ast\sim\mu^{1/2}\) is recovered as the leading-order term.

The original variables are reconstructed through a smooth embedding reversing the normal form transformation,
\begin{equation}
x-x_\ast(\mu)=\Phi(z,\bar z,\mu)\,,
\end{equation}
where \(\Phi\) admits a local power series expansion as a near-identity analytic transformation \cite{guck83,craw91}.
This yields the observable expansions used below.


\section{Expansion of the excess observable and cancellation of odd orders}
\label{supp:odd}

\subsection{Fourier representation}

We derive the integer-power expansion of the excess observable
\begin{equation}
\Delta A(\mu)\equiv \langle A\rangle-A_\ast(\mu)\,,
\qquad
A_\ast(\mu)\equiv A(x_\ast(\mu))\,,
\end{equation}
where \(\langle\cdot\rangle\) denotes averaging over one period of the stable limit cycle.

The key geometric consequence of Eq.~(\ref{eq:app_polar_nf2}) is that on the stable limit cycle the amplitude is constant,
\begin{equation}
r=r_\ast(\mu)\,,
\end{equation}
while the phase variable \(\theta\) evolves periodically. After radial relaxation, the attractor therefore has only one remaining dynamical degree of freedom: the phase.

Hence, in the long-time limit, the trajectory of the limit cycle may be written as
\begin{equation}
x(\theta,\mu)=x_\ast(\mu)+\delta x(\theta,\mu)\,,
\end{equation}
where \(\delta x\) is smooth and \(2\pi\)-periodic in \(\theta\).

Because the inverse center-manifold embedding \(\Phi(z,\bar z,\mu)\) is smooth and
\begin{equation}
z=r_\ast(\mu)e^{i\theta}\,,
\qquad
\bar z=r_\ast(\mu)e^{-i\theta}\,,
\end{equation}
the oscillatory displacement admits the expansion
\begin{equation}
\small
\delta x(\theta,\mu)
=
\Phi(z,\bar z,\mu)
=
\sum_{n\ge1}\sum_{\tilde n=0}^{n}
\phi_{n-\tilde n,\tilde n}(\mu)\,
e^{i(n-2\tilde n)\theta}\,
r_\ast(\mu)^n \,,
\label{eq:app:dx_z}
\end{equation}
where \(\phi(\mu)_{l,k}\) denote the series coefficients of \(\Phi(z,\bar z,\mu)\) in powers of \(z^l\bar z^k\).

Equivalently, this defines the amplitude-resolved Fourier representation
\begin{equation}
\delta x(\theta,\mu) \equiv \sum_{n\geq1} X_n(\theta,\mu)r_\ast(\mu)^n\,,
\label{eq:app_Xk_exp}
\end{equation}
where each \(X_n\) is a finite trigonometric polynomial with coefficients smooth in \(\mu\). The leading orders are for instance
\begin{align}
X_1(\theta,\mu)
&=
u(\mu)\cos\theta+v(\mu)\sin\theta\,,
\\
X_2(\theta,\mu)
&=
X_{2,0}(\mu)+u_2(\mu)\cos2\theta+v_2(\mu)\sin2\theta\,,
\end{align}
describing the fundamental harmonic and its lowest nonlinear deformation.


\subsection{Expansion of the excess observable}

Let \(A(x)\) be a smooth scalar observable for which local expansions are well defined.
Expanding around the smooth stationary branch gives
\begin{equation}
\begin{split}
A(x_\ast+\delta x)-A(x_\ast)
&=
\sum_{m\ge1}\frac{1}{m!}
D^mA(x_\ast(\mu))
[\delta x,\ldots,\delta x]
\\
&=
\nabla A(x_\ast)\!\cdot\!\delta x
+
\frac12\delta x^\top H_A(x_\ast)\delta x
+\ldots \,,
\end{split}
\label{eq:app_taylor_delta}
\end{equation}
where \(D^mA\) denotes derivative tensor of order \(m\) and \(H_A\) the Hessian of \(A\).

Substituting Eq.~(\ref{eq:app:dx_z}) into Eq.~(\ref{eq:app_taylor_delta}) yields
\begin{widetext}
\begin{equation}
\begin{split}
A(x_\ast+\delta x)-A(x_\ast)
&=
\sum_{m\ge1}\frac{1}{m!}
\Bigg\{
\sum_{n_1,\ldots,n_m\ge1}
\sum_{\tilde n_1=0}^{n_1}\cdots
\sum_{\tilde n_m=0}^{n_m}
D^mA(x_\ast(\mu))
[
\phi_{n_1-\tilde n_1,\tilde n_1},
\ldots,
\phi_{n_m-\tilde n_m,\tilde n_m}
]
e^{\,i\theta\sum_{l=1}^{m}(n_l-2\tilde n_l)}
\,r_\ast(\mu)^{\sum_{l=1}^{m}n_l}
\Bigg\}\,.
\end{split}
\label{eq:deltaA_long}
\end{equation}
\end{widetext}

\subsection{Phase averaging and even powers}

Taking the phase average affects only the exponential factor:
\begin{equation}
\left\langle
e^{\,i\theta\sum_{l=1}^{m}(n_l-2\tilde n_l)}
\right\rangle
=
\delta_{\sum_{l=1}^{m}n_l-\;2\sum_{l=1}^{m}\tilde n_l,0}\,,
\end{equation}
where \(\delta_{\bullet,\bullet}\) denotes the Kronecker delta.

Hence only terms with vanishing total Fourier index survive. Applying this constraint to Eq.~(\ref{eq:deltaA_long}) gives
\begin{widetext}
\begin{equation}
\begin{split}
\Delta A(\mu)
&=
\sum_{m\ge1}\frac{1}{m!}
\Bigg\{
\sum_{n_1,\ldots,n_m\ge1}
\sum_{\tilde n_1=0}^{n_1}\cdots
\sum_{\tilde n_m=0}^{n_m}
D^mA(x_\ast(\mu))
[
\phi_{n_1-\tilde n_1,\tilde n_1},
\ldots,
\phi_{n_m-\tilde n_m,\tilde n_m}
]
r_\ast(\mu)^{2\sum_{l=1}^{m}\tilde n_l}
\Bigg\}\,.
\end{split}
\label{eq:app:delta_A_shorter}
\end{equation}\end{widetext}
Thus the limit-cycle amplitude \(r_\ast(\mu)\) appears only with even integer powers, \(r_\ast(\mu)^{2\sum_{l=1}^{m}\tilde n_l}\).

Rearranging the sums in Eq.~(\ref{eq:app:delta_A_shorter}) therefore yields
\begin{equation}
\Delta A(\mu)
=
\sum_{j\ge1}b_j(\mu)\,r_\ast(\mu)^{2j}\,,
\label{eq:app_even_r}
\end{equation}
where each coefficient \(b_j(\mu)\) is smooth in \(\mu\) and collects all symmetry-allowed couplings between the local derivatives of the observable and the Fourier components of the emergent waveform at total order \(r_\ast^{2j}\).

Using \(r_\ast(\mu)^2=s(\mu)\), and both \(b_j(\mu)\) and \(s(\mu)\) admit one-sided power series in \(\mu\) with \(s(\mu)\) starting at the linear order, compare Eq.~(\ref{eq:app_smooth_s}), their products are ordinary integer-power series.
Hence, it follows that
\begin{equation}
\Delta A(\mu)
=
\sum_{m\ge1}a_m\mu^m\,,
\qquad
\mu>0\,,
\end{equation}
which proves the expansion stated in the main text.

Thus higher-order corrections to the cycle amplitude and smooth waveform deformations modify only the coefficients \(a_m\); they do not generate fractional powers or divergences for a generic smooth supercritical Hopf bifurcation.


\section{Practical evaluation of the leading coefficient}
\label{supp:practical_hierarchy}

We derive an explicit expression for the leading coefficient \(a_1\) in
\begin{equation}
\Delta A(\mu)=a_1\mu+\mathcal O(\mu^2)\,,
\qquad \mu>0\,,
\end{equation}
using only local quantities at the Hopf bifurcation introduced above, thereby connecting \(a_1\) to standard quantities from Hopf normal-form theory. For transparency, we present the derivation for a two-dimensional system. The generalization to higher dimensions is discussed in Sec.~\ref{sec:supp:higherDim}.

\subsection{Amplitude prefactor}

The Hopf amplitude equation reads \cite{guck83,craw91,wigg03}
\begin{equation}
\dot r=\kappa \mu r+l_1 r^3+\ldots\,,
\end{equation}
where \(\kappa\equiv \alpha'(0)\), with \(\alpha(\mu)\) defined in Eq.~(\ref{eq:app:eigenvalues}), and \(l_1 \equiv \operatorname{Re}\ell_1(0)\) is the first Lyapunov coefficient. The latter can be obtained as follows.

Let \(x_c=x_\ast(0)\) denote the critical fixed point.
After shifting \(x_c\) to the origin and linearly transforming \(L(0)\), compare Eq.~(\ref{eq:app:Jacobian}), into the canonical rotation block, the dynamics take the form
\begin{align}
\dot \xi &= -\omega_c \eta + F(\xi,\eta)\,,\\
\dot \eta &= \phantom{-}\omega_c \xi + G(\xi,\eta)\,,
\end{align}
where \(F\) and \(G\) contain only nonlinear terms of order at least two, and
\(\omega_c\equiv \omega(0)\), with \(\omega(\mu)\) defined in Eq.~(\ref{eq:app:eigenvalues}).

For planar systems, the first Lyapunov coefficient is \cite{guck83,wigg03}
\begin{align}
l_1
&=
\frac{1}{16}
\Bigl(
F_{\xi\xi\xi}+F_{\xi\eta\eta}+G_{\xi\xi\eta}+G_{\eta\eta\eta}
\Bigr)
\nonumber\\
&\quad+
\frac{1}{16\omega_c}
\Bigl[
F_{\xi\eta}(F_{\xi\xi}+F_{\eta\eta})
-
G_{\xi\eta}(G_{\xi\xi}+G_{\eta\eta})
\nonumber\\
&\qquad\qquad
-
F_{\xi\xi}G_{\xi\xi}
+
F_{\eta\eta}G_{\eta\eta}
\Bigr]\,,
\label{eq:app_explicit_l1}
\end{align}
with all derivatives evaluated at \((\xi,\eta)=(0,0)\).

For a supercritical Hopf bifurcation with stable oscillations for \(\mu>0\),
\begin{equation}
\kappa>0,
\qquad
l_1<0,
\end{equation}
so that the cycle amplitude satisfies
\begin{equation}
r_\ast^2=c_1\mu+\mathcal O(\mu^2)\,,
\qquad
c_1=-\frac{\kappa}{l_1}\,.
\label{eq:app_unified_c1}
\end{equation}

\subsection{Leading waveform coefficients}

At the bifurcation, the first Fourier component of the cycle displacement can be written as
\begin{equation}
X_1(\theta,0)=X_{1,c}\cos\theta+X_{1,s}\sin\theta\,,
\label{eq:app_X1_mode}
\end{equation}
where \(X_{1,c},X_{1,s}\in\mathbb R^2\) span the Hopf eigenspace of \(L(0)\). Equivalently, they are the real and imaginary parts of the critical complex eigenvector.

The second Fourier component has the structure
\begin{equation}
X_2(\theta,0)=X_{2,0}+X_{2,c}\cos2\theta+X_{2,s}\sin2\theta\,.
\end{equation}
Its phase-independent part \(X_{2,0}\) represents the quadratic mean deformation of the emerging cycle.

Let
\begin{equation}
F_2=D_x^2f(x_c,0)
\end{equation}
denote the quadratic derivative tensor with respect to the state variables. Expanding
\begin{equation}
\delta x=x-x_c=r_\ast X_1+r_\ast^2X_2+\mathcal O(r_\ast^3)
\label{eq:app:delta_x}
\end{equation}
in the shifted dynamics
\begin{equation}
\dot{\delta x}=L(0)\delta x+\frac12F_2(\delta x,\delta x)+\ldots
\end{equation}
gives at order \(r^2\),
\begin{equation}
\omega_c\partial_\theta X_2
=
L(0)X_2+\frac12F_2(X_1,X_1)\,.
\end{equation}

Averaging over one period removes the derivative term and all oscillatory modes, yielding
\begin{equation}
0=
L(0)X_{2,0}
+
\frac12\left\langle F_2(X_1,X_1)\right\rangle\,.
\end{equation}
For a generic Hopf bifurcation, \(L(0)\) has full rank and is therefore invertible. Hence, evaluating the phase-average yields
\begin{equation}
X_{2,0}
=
-\frac14L(0)^{-1}
\left[
F_2(X_{1,c},X_{1,c})
+
F_2(X_{1,s},X_{1,s})
\right]\,.
\label{eq:app_unified_X20}
\end{equation}

\subsection{Leading observable coefficient}

Let \(\nabla A\) and \(H_A\) denote the gradient and Hessian of the observable evaluated at \(x_c=x_\ast(0)\). Expanding \(A\) around the smoothly continued stationary branch \(x_\ast(\mu)\), and then evaluating the leading coefficient at \(\mu=0\), gives
\begin{equation}
A(x_\ast+\delta x)-A(x_\ast)
=
\nabla A\cdot\delta x
+
\frac12\delta x^\top H_A\delta x+\ldots\,,
\end{equation}
where the derivatives are evaluated at \(x_c\) in the leading-order coefficient. Averaging with Eq.~(\ref{eq:app:delta_x}) shows that
\begin{equation}
\Delta A=b_1(0)\,r_\ast^2+\mathcal O(r_\ast^4)\,,
\end{equation}
where
\begin{equation}
b_1(0)=
\nabla A\cdot X_{2,0}
+
\frac14
\left(
X_{1,c}^\top H_A X_{1,c}
+
X_{1,s}^\top H_A X_{1,s}
\right)\,.
\label{eq:app_unified_b1}
\end{equation}

Using Eq.~(\ref{eq:app_unified_c1}) then gives
\begin{equation}
a_1=b_1(0)c_1\,,
\end{equation}
hence
\begin{equation}
a_1=
\left[
\nabla A\cdot X_{2,0}
+
\frac14
\left(
X_{1,c}^\top H_A X_{1,c}
+
X_{1,s}^\top H_A X_{1,s}
\right)
\right]
\left(-\frac{\kappa}{l_1}\right)\,.
\label{eq:app_unified_a1}
\end{equation}

Eq.~(\ref{eq:app_unified_a1}) provides the explicit leading-order realization of the general coupling structure derived in Sec.~\ref{supp:odd}. It separates the leading kink amplitude into a dynamical factor \(-\kappa/l_1\), which controls the growth of the squared cycle amplitude, and a geometric projection factor, which encodes how the observable couples to the quadratic mean deformation and fundamental harmonic structure of the emergent waveform. Vanishing of this projection suppresses the generic kink and shifts the Hopf singularity to higher Ehrenfest order.

\subsection{Higher dimensions}
\label{sec:supp:higherDim}
The derivation above was written explicitly for planar systems, where the Hopf eigenspace spans the full state space. For dimensions \(n>2\), the same structure follows after standard center-manifold reduction to the critical two-dimensional subspace. The coefficient formula therefore remains valid in arbitrary finite dimension: only the explicit expression for \(l_1\) is replaced by the corresponding higher-dimensional Hopf formulas \cite{kuzn23}, while the quadratic mean shift is obtained from the full linearization and nonlinear tensors in the ambient space.


\section{Numerical evaluation of the leading Hopf coefficient for the reversible Brusselator}

In this section we summarize a practical numerical procedure for evaluating the leading coefficient \(g_1\) in the reversible Brusselator expansion
\begin{equation}
\Delta \mathcal G(\mu)=g_1\mu+\mathcal O(\mu^2)\,,
\end{equation}
where \(\Delta\mathcal G\) is the excess semi-grand Gibbs free energy and
\begin{equation}
\mu=\mathcal A-\mathcal A_c
\end{equation}
denotes the distance from the Hopf bifurcation.

The calculation provides an independent benchmark for trajectory-based estimates obtained from direct time averaging of the nonlinear dynamics.


\subsection{Model}
The chemical reaction scheme is given by 
\begin{equation}
A \crn{-1}{1} X\,,\qquad
3X \crn{2}{-2} 2X+Y\,,\qquad
Y \crn{3}{-3} B\,,
\label{eq:supp:CRNBruss}
\end{equation}
where \(X\) and \(Y\) label the internal species and \(A\) and \(B\) the chemostatted species with constant concentrations denoted \([A]\) and \([B]\), respectively. Under mass-action kinetics \cite{gasp04}, the deterministic reaction rate equations read
\begin{equation}
\begin{split}
\dot x \equiv \begin{pmatrix}
\dot x_1 \\ \dot x_2 \end{pmatrix}    = \begin{pmatrix}J_1+J_2\\
-J_2+J_3\end{pmatrix} \equiv f(x,\mathcal A)\,,
\end{split}
\end{equation}
with reaction fluxes
\begin{align}
J_1 &= k_{1}[A]-k_{-1}x_1\,,\\
J_2 &= k_{2}x_1^2x_2-k_{-2}x_1^3\,,\\
J_3 &= k_{3}[B]-k_{-3}x_2\,,
\end{align}
where, for convenience, we labeled the concentrations of the internal species \(X\) and \(Y\) as \(x_1\) and \(x_2\), respectively.
The deterministic dynamics undergoes a Hopf bifurcation while varying the affinity
\begin{equation}
\mathcal A=
\ln\frac{[B]k_3k_2k_{-1}}{[A]k_1k_{-2}k_{-3}}\,,\label{eq:supp:LDB}
\end{equation}
so that varying \(\mathcal A\) corresponds to varying the reverse rate \(k_{-3}\), while keeping the remaining kinetic constants fixed. The parameter set used in the main text reads (in arbitrary units)
\begin{equation}
\begin{split}
k_1&=0.1\,,\quad
k_{-1}=1\,,\quad
k_2=1\,,\quad
k_{-2}=1\,,\quad\\
k_3&=0.1\,,\quad
[A]=1\,,\quad
[B]=3\,.\end{split} \label{eq:supp:brusselRates}
\end{equation}


\subsection{Hopf bifurcation and stationary branch}

For each value of \(\mathcal A\), the stationary branch
\begin{equation}
x_\ast(\mathcal A)=\bigl(x_{1,\ast},x_{2,\ast}\bigr)
\end{equation}
is obtained numerically by solving
\begin{equation}
f(x,\mathcal A)=0\label{eq:supp:stationary}\,.
\end{equation}

The Hopf point \((x_c,\mathcal A_c)\) is then determined from the stationary conditions together with the planar Hopf criterion
\begin{equation}
\operatorname{tr}L(\mathcal A_c)=0\,,\label{eq:supp:traceCond}
\qquad
\det L(\mathcal A_c)>0\,,
\end{equation}
where
\begin{equation}
L(\mathcal A)\equiv D_x f(x_\ast(\mathcal A),\mathcal A)
\end{equation}
is the Jacobian evaluated on the stationary branch.

For the above parameters, Eq. (\ref{eq:supp:brusselRates}), solving Eq. (\ref{eq:supp:stationary}) and (\ref{eq:supp:traceCond}) simultaneously yields
\begin{align}
\mathcal A_c &\simeq 3.93034537\,,\\
x_c=x_\ast(\mathcal A_c)
&\simeq
(0.24302234,\;2.66467295)\,.
\end{align}

Introducing the shifted control parameter
\begin{equation}
\mu=\mathcal A-\mathcal A_c\,,
\end{equation}
the Hopf bifurcation is located at \(\mu=0\).


\subsection{Semi-grand Gibbs free energy}

Following the framework of Ref.~\cite{rao18} for thermodynamics of chemical reaction networks, the reversible Brusselator admits a semi-grand Gibbs free energy acting as a Lyapunov-like nonequilibrium potential under deterministic mass-action kinetics,
\begin{equation}\small\begin{split}
\mathcal G(x_1,x_2;\mathcal A)
&=
x_1\ln\!\frac{k_{-2}k_{-3}x_1}{k_2k_3}
+x_2\ln\!\frac{k_{-3}x_2}{k_3}
-x_1-x_2-[A]-[B]
\\
&\quad
-\bigl([A]+x_1+x_2\bigr)\ln[B]
+[A]\ln\!\frac{[A]k_1k_{-2}k_{-3}}{k_{-1}k_2k_3}\,,
\end{split}\end{equation}
where \(k_{-3}\) is understood as a function of \(\mathcal A\) through Eq. (\ref{eq:supp:LDB}).

Its excess value along the stable limit cycle is
\begin{equation}
\Delta\mathcal G(\mu)
=
\langle \mathcal G\rangle
-
\mathcal G(x_\ast(\mu);\mathcal A_c+\mu)\,.
\end{equation}


\subsection{Local derivatives at the Hopf point}

At \((x_c,\mathcal A_c)\), the tensors entering the analysis as introduced in the main text are evaluated numerically:
\begin{align}
L(0) &= D_x f(x_c,\mathcal A_c)\,,\\
F_2 &= D_x^2 f(x_c,\mathcal A_c)\,,\\
F_3 &= D_x^3 f(x_c,\mathcal A_c)\,.
\end{align}

The critical eigenvalue pair of \(L(0)\) is
\begin{equation}
\lambda_\pm=\pm i\omega_c\,,
\end{equation}
with
\begin{equation}
\omega_c\simeq 0.22827644\,,
\end{equation}
and corresponding real eigendirections \(X_{1,c},X_{1,s}\).


\subsection{Evaluation of \(\kappa\)}

The coefficient
\begin{equation}
\kappa=\alpha'(0)
\end{equation}
is obtained numerically from the real part of the complex conjugate eigenvalues along the stationary branch:
\begin{equation}
\lambda_\pm(\mu)=\alpha(\mu)\pm i\omega(\mu)\,.
\end{equation}

Using a centered finite difference,
\begin{equation}
\kappa \approx
\frac{\alpha(+\delta)-\alpha(-\delta)}{2\delta}\,,
\end{equation}
with \(\delta\ll1\), yielding
\begin{equation}
\kappa\simeq 0.24422370\,.
\end{equation}


\subsection{Evaluation of \(l_1\)}

The first Lyapunov coefficient \(l_1\) is computed from the local derivatives of the planar vector field using the standard explicit Hopf formula, Eq. (\ref{eq:app_explicit_l1}). For the present parameters,
\begin{equation}
l_1\simeq -0.0834277.
\end{equation}

This determines the leading cycle amplitude
\begin{equation}
r_\ast^2=c_1\mu+\mathcal O(\mu^2)\,,
\qquad
c_1=-\frac{\kappa}{l_1}\simeq 2.92737\,.
\end{equation}


\subsection{Observable projection}

At the Hopf point we evaluate
\begin{equation}
\nabla_x \mathcal G(x_c,\mathcal A_c)\,,
\qquad
H_x\mathcal G(x_c,\mathcal A_c)\,,
\end{equation}
with derivatives taken with respect to the state variables \(x=(x_1,x_2)\).

The second-order mean displacement is
\begin{equation}\begin{split}
X_{2,0}
&=
-\frac14L(0)^{-1}
\left[
F_2(X_{1,c},X_{1,c})
+
F_2(X_{1,s},X_{1,s})
\right] \\ &\simeq (0.027362685959,\; -0.464477604683)\,.\end{split}
\end{equation}

The two contributions to \(b_1(0)\) are
\begin{align}
\nabla_x\mathcal G\cdot X_{2,0}
&\simeq 0.217585\,,\\
\frac14
\left(
X_{1,c}^\top H_x\mathcal G\,X_{1,c}
+
X_{1,s}^\top H_x\mathcal G\,X_{1,s}
\right)
&\simeq 0.140730\,.
\end{align}

Hence
\begin{equation}
b_1(0)\simeq 0.358315\,.
\end{equation}

The leading coefficient then follows from
\begin{equation}
g_1\equiv b_1(0)\left(-\frac{\kappa}{l_1}\right)
\simeq 1.048921240522\,.
\end{equation}


\subsection{Entropy production rate / chemical work rate}

For an open chemical reaction network, the semi-grand Gibbs free energy satisfies the first-law-like balance relation \cite{rao16,rao18}
\begin{equation}
    d_t\mathcal G
    =
    -\dot\Sigma
    +
    \dot W\,.
\end{equation}
Since \(\mathcal G\) is a state function, its time derivative averages to zero both at stationary states and on stable periodic orbits,
\begin{equation}
\left\langle d_t \mathcal G\right\rangle =0\,.
\end{equation}
It therefore follows that the time-averaged entropy production rate equals the time-averaged chemical work rate,
\begin{equation}
\langle \dot\Sigma\rangle
=
\langle \dot W\rangle\,.
\end{equation}

The chemical work rate consistent with the definition of \( \mathcal G\) is
\begin{equation}
\dot W=-\mathcal A\,J_1\,,
\end{equation}
where \(\mathcal A\) is the chemical affinity defined in Eq.~(\ref{eq:supp:LDB}) and \(J_1\) denotes the exchange flux associated with the chemostatted species \(A\). Along the stable periodic orbit, the reaction currents are time dependent and generally distinct. However, because the limit cycle constitutes a periodic steady state, the time-averaged reaction rate equations imply
\begin{equation}
\langle \dot x_1\rangle
=
\langle J_1+J_2\rangle
=0\,,
\end{equation}
and
\begin{equation}
\langle \dot x_2\rangle
=
\langle -J_2+J_3\rangle
=0\,.
\end{equation}
Hence the time-averaged cycle currents satisfy,
\begin{equation}
\langle J_1\rangle
=
-\langle J_2\rangle
=
-\langle J_3\rangle\,.
\end{equation}

We therefore consider the excess work rate
\begin{equation}
\Delta \dot W(\mu)
=
\langle \dot W\rangle
-
\dot W(x_\ast(\mu))\,.
\end{equation}
Since the averaged entropy production rate and averaged work rate coincide, one equivalently has
\begin{equation}
\Delta \dot\Sigma(\mu)=\Delta \dot W(\mu)\,.
\end{equation}

Applying the leading-coefficient formula Eq.~(\ref{eq:app_unified_a1}) numerically yields
\begin{equation}
\Delta \dot W
=
0.314823300822\,\mu
+
\mathcal O(\mu^2)\,.
\end{equation}
Hence
\begin{equation}
n_\ast^{\dot\Sigma}=n_\ast^{\dot W}=1\,.
\end{equation}

Thus, for the reversible Brusselator, both the semi-grand Gibbs free energy and the entropy production rate realize the generic Hopf kink class. This contrasts with the three-stage CMOS ring oscillator, where the linear entropy-production contribution vanishes due to symmetry-induced cancellation, shifting the singularity to \(n_\ast^{\dot\Sigma}=2\).

\begin{figure}[t]
\centering
\includegraphics[width=\figwidth]{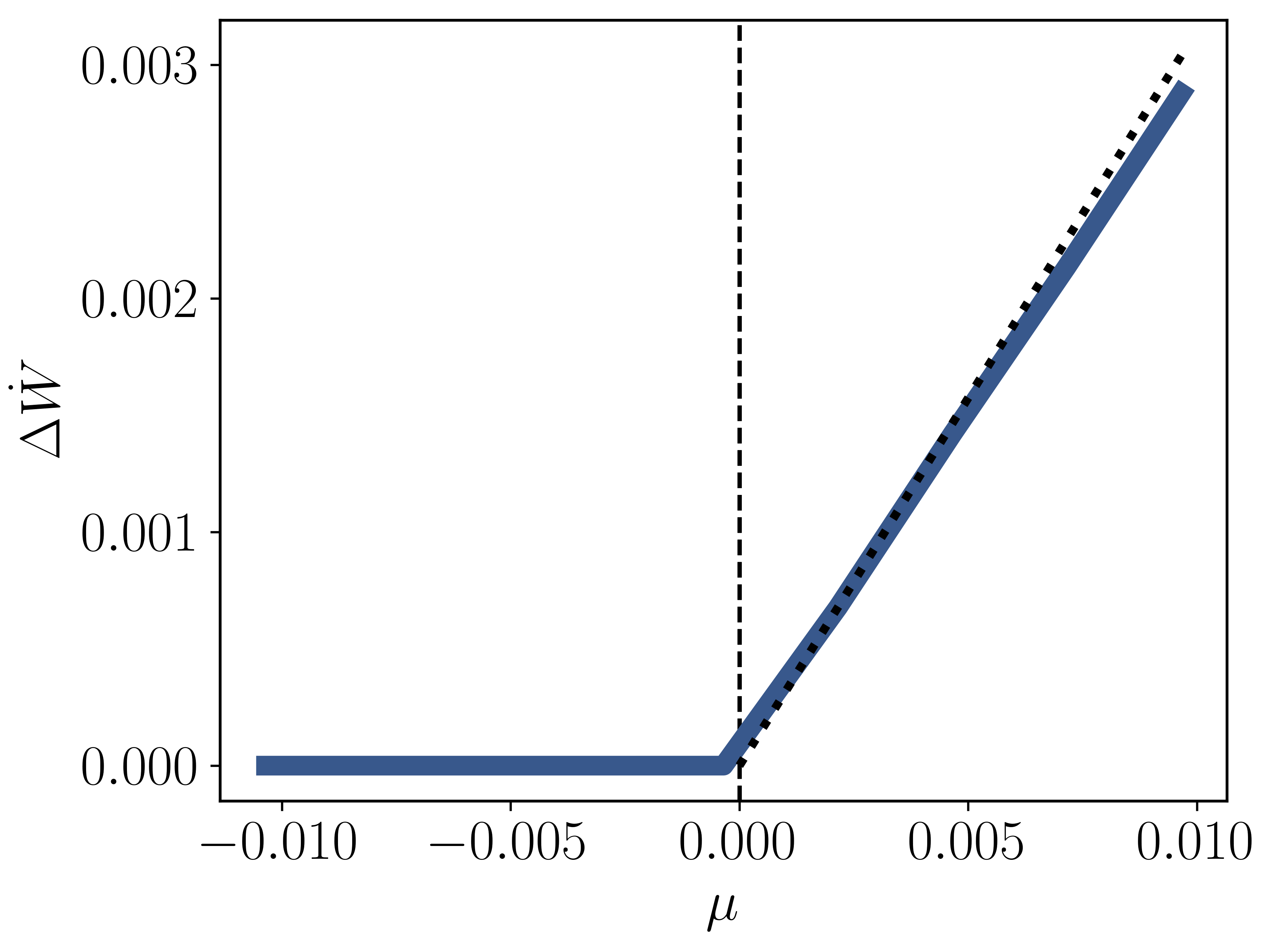}
\caption{
Excess chemical work rate
\(
\Delta \dot W
=
\langle \dot W\rangle
-
\dot W_\ast
\)
for the reversible Brusselator near the Hopf bifurcation as function of the distance
\(
\mu=\mathcal A-\mathcal A_c
\)
from criticality.
The solid curve shows direct numerical time averaging of the nonlinear dynamics, while the dotted line indicates the local Hopf prediction
\(
\Delta\dot W = w_1 \mu
\)
with
\(
w_1 = 0.314823300822
\),
obtained from the leading-order coefficient formula derived in Sec.~\ref{supp:practical_hierarchy}.
The agreement confirms the predicted generic kink singularity
\(
n_\ast^{\dot W}=1
\).
}
\label{fig:brusselator_work}
\end{figure}


\section{Thermodynamic structure of the CMOS ring oscillator}
\label{supp:CMOS}

In this section, we briefly summarize the thermodynamic structure underlying the CMOS ring oscillator discussed in the main text, following Ref.~\cite{gopa25}.

The \(N\)-stage ring oscillator consists of a cyclic chain of coupled CMOS inverters driven by a constant voltage supply. After suitable nondimensionalization, the deterministic voltage dynamics takes the form
\begin{equation}
\dot x_i=
2\left[
\sinh(x_{i-1}-x_i)-e^{a}\sinh(x_{i-1})
\right] \equiv f_i(x,a)\,,
\label{eq:CMOSdriftSM}
\end{equation}
\(i=1,\dots,N,\) with cyclic indexing. The parameter \(a\) quantifies the nonequilibrium electrical drive relative to thermal fluctuations.

The vector-field \( f(x,a) \), Eq.~(\ref{eq:CMOSdriftSM}), originates from the macroscopic limit of a microscopic description of the transition currents through the inverters \cite{gopa25,frei21b}.
The macroscopic rates are (in dimensionless quantities) given by
\begin{equation}
\begin{split}    w_{i+}^p(x) &\equiv e^{a - x_{i-1}}\equiv w_{i+}^n(-x)\,,\\w_{i-}^p(x) &\equiv e^{x_i - x_{i-1}}\equiv w_{i-}^n(-x)\,.
\end{split}\end{equation}
Taking into account only the restricted transitions due to a single charge transfer, the vector field reads
\begin{equation}
f_i(x)=\sum_{\rho=n,p} \eta_\rho [ w_{i+}^\rho(x)- w_{i-}^\rho(x)]\,,
\end{equation}
where  \(\eta_p \equiv + 1\) and \(\eta_n \equiv -1\).
The underlying microscopic description satisfies a local detailed balance condition, thus, in the macroscopic limit, a physically meaningful entropy production rate arises as
\begin{equation}
\dot \Sigma
=
\sum_i\sum_{\rho=n,p}
(w_{i+}^\rho(x) - w_{i-}^\rho(x))
\ln\frac{w_{i+}^\rho(x)}{w_{i-}^\rho(x)},
\label{eq:EPRgeneralCMOS}
\end{equation}
which has the standard current--affinity structure of stochastic thermodynamics. Evaluating Eq.~(\ref{eq:EPRgeneralCMOS}) for the CMOS ring oscillator yields
\begin{equation}
\dot \Sigma
=
2a
\sum_i
\left[
e^a\cosh(x_{i-1})
-
\cosh(x_i-x_{i-1})
\right].
\label{eq:EPRCMOSSM}
\end{equation}

The stationary operating point loses stability through a supercritical Hopf bifurcation at
\begin{equation}
a_c=\log\!\left(1+\sec\frac{\pi}{N}\right).
\end{equation}
Introducing the distance from the bifurcation,
\begin{equation}
\mu=a-a_c,
\end{equation}
the oscillation amplitude is given in Eq.~(36) of Ref.~\cite{gopa25},
\begin{equation}
    r_\ast^2 \simeq \frac{N^2}{2 \left[\cos^2\left(\frac{\pi}{2N}\right)-\frac 14\right]}\mu\label{eq:supp:CMOSr}
\end{equation}
while Eq.~(43) gives the excess entropy production rate
\begin{equation}
    \Delta \dot \Sigma \simeq \frac{a}{N}\left[ e^a - 4 \cos^2\left(\frac{\pi}{2N} \right)\right]r_\ast^2\,.\label{eq:supp:CMOSepr}
\end{equation}
Plugging Eq. (\ref{eq:supp:CMOSr}) into Eq. (\ref{eq:supp:CMOSepr}) and expanding in \(\mu \ll 1\), yields
\begin{equation}\small\begin{split}
\Delta \dot \Sigma
&=
-4N\ln\!\left(1+\sec\frac{\pi}{N}\right)
\left[
1-\frac{1}{
\cos(\pi/N)\left(1+2\cos(\pi/N)\right)}
\right]\mu + \mathcal O(\mu^2) \\ &\equiv s_1^{(N)} \mu + \mathcal O(\mu^2).
\end{split}\end{equation}
For \( N =3 \), the prefactor vanishes, \(s_1^{(N = 3)} = 0\), leading to \(n_\ast^{\dot \Sigma} = 2\), while for \(N > 3\), \(s_1^{(N)}\) remains finite, thus, places the entropy production rate of the CMOS oscillator in the singularity class \(n_\ast^{\dot \Sigma} = 1\).

Using the leading-order coefficient formula, Eq.~(\ref{eq:app_unified_a1}), we gain geometrical insight into the cancellation of the first-order contribution for the \(N=3\) oscillator. The vector field \(f(x,a)\) is antisymmetric, implying that the quadratic contribution to the normal form vanishes, \(F_2=0\), and therefore the associated center-manifold correction satisfies \(X_{20}=0\). Consequently, the gradient contribution to the leading-order coefficient vanishes, leaving only the contribution due to the Hessian of the entropy production rate. For \(N=3\), the resulting quadratic form vanishes along the critical Hopf mode, suppressing the generic linear singularity and leading to the quadratic onset \(n_\ast^{\dot\Sigma}=2\).


\section{ENSO recharge--discharge model and evaluation of \(a_1\)}
\label{sec:SM_ENSO}

We apply the general leading-coefficient formula of the main text to the recharge--discharge oscillator model of ENSO,
\begin{align}
\dot T &= C T + D h - \varepsilon T^3\,,
\label{eq:SM_ENSO_T}\\
\dot h &= -E T - R_h h\,.
\label{eq:SM_ENSO_h}
\end{align}
The state vector is \(x=(T,h)\), and we consider the observable
\begin{equation}
A(x)=T^2.
\end{equation}

The stationary branch is
\begin{equation}
x_\ast(C)=(0,0)\,,
\end{equation}
which remains smooth through the bifurcation. Linearizing around this branch gives
\begin{equation}
L(C)=
\begin{pmatrix}
C & D\\
-E & -R_h
\end{pmatrix}\,.
\end{equation}
The location of the Hopf bifurcation is determined by
\begin{equation}
\operatorname{tr}L=0\,,
\qquad
\det L>0\,,
\end{equation}
hence
\begin{equation}
C_c=R_h\,,
\qquad
\omega_c^2=DE-R_h^2>0\,.
\end{equation}
We define the distance from the bifurcation as
\begin{equation}
\mu=C-C_c=C-R_h\,.
\end{equation}

\subsection{Ingredients of the \(a_1\) formula}

At the onset of oscillations, the leading-order periodic orbit lies in the critical eigenspace.
Choosing the phase origin and normalizing the temperature component,
we write
\begin{equation}
T(t)=r\cos(\omega_c t)+O(r^2)\,,
\end{equation}
while the thermocline component is determined consistently from the linearized dynamics as follows.
Using the linearized equation
\begin{equation}
\dot T=R_h T+D h\,,
\end{equation}
this implies
\begin{equation}
h(t)
=
-\frac{R_h}{D}r\cos(\omega_c t)
-\frac{\omega_c}{D}r\sin(\omega_c t)
+\mathcal O(r^2)\,.
\end{equation}
Thus the leading Hopf mode has the form
\begin{equation}
X_1(\theta)=X_{1,c}\cos\theta+X_{1,s}\sin\theta,
\end{equation}
with
\begin{equation}
X_{1,c}=
\begin{pmatrix}
1\\
-R_h/D
\end{pmatrix}\,,
\qquad
X_{1,s}=
\begin{pmatrix}
0\\
-\omega_c/D
\end{pmatrix}\,.
\end{equation}

For the observable \(A(T,h)=T^2\), evaluated at the Hopf point \(x_c=(0,0)\), one has
\begin{equation}
\nabla A(x_c)=
\begin{pmatrix}
0\\
0
\end{pmatrix}\,,
\qquad
H_A(x_c)=
\begin{pmatrix}
2 & 0\\
0 & 0
\end{pmatrix}\,.
\end{equation}
Therefore the mean-shift contribution in the general formula vanishes,
\begin{equation}
\nabla A\cdot X_{2,0}=0\,,
\end{equation}
and
\begin{align}
b_1(0)
&=
\frac14
\left(
X_{1,c}^{\top}H_A X_{1,c}
+
X_{1,s}^{\top}H_A X_{1,s}
\right)
\nonumber\\
&=
\frac14(2+0)
=
\frac12\,.
\end{align}

\subsection{Amplitude prefactor from the Hopf normal form}

With \(C=R_h+\mu\), the Jacobian along the stationary branch \(x_\ast=(0,0)\) is
\begin{equation}
L(\mu)=
\begin{pmatrix}
R_h+\mu & D\\
-E & -R_h
\end{pmatrix}\,.
\end{equation}
Its trace and determinant are
\begin{equation}
\operatorname{tr}L(\mu)=\mu\,,
\qquad
\det L(\mu)=DE-R_h^2-R_h\mu \,.
\end{equation}
The eigenvalues are therefore
\begin{equation}\begin{split}
\lambda_\pm(\mu) &=\frac{\operatorname{tr}L(\mu)}{2}
\pm
\sqrt{
\left(\frac{\operatorname{tr}L(\mu)}{2}\right)^2
-
\det L(\mu)
}
\\ &=
\frac{\mu}{2}
\pm
\sqrt{
\frac{\mu^2}{4}
-
\bigl(DE-R_h^2-R_h\mu\bigr)
}\,.\end{split}
\end{equation}
At the Hopf point \(\mu=0\),
\begin{equation}
\lambda_\pm(0)=\pm i\omega_c\,,
\qquad
\omega_c^2=DE-R_h^2>0\,.
\end{equation}
Thus the real part satisfies
\begin{equation}
\alpha(\mu)=\frac{\mu}{2}+\mathcal O(\mu^2)\,,
\end{equation}
and hence
\begin{equation}
\kappa=\alpha'(0)=\frac12\,.
\end{equation}

To evaluate \(l_1\), we transform the linearized dynamics at \(\mu=0\) into canonical rotation form. Let
\begin{equation}
\xi=T\,,
\qquad
\eta=-\frac{R_h T+D h}{\omega_c}\,.
\end{equation}
Then, at \(\mu=0\),
\begin{align}
\dot \xi
&=
-\omega_c\eta
-\varepsilon \xi^3\,,
\\
\dot \eta
&=
\omega_c\xi
+
\frac{R_h\varepsilon}{\omega_c}\xi^3\,.
\end{align}
Thus, in the notation of the explicit planar Hopf formula,
\begin{align}
\dot \xi &= -\omega_c\eta + F(\xi,\eta)\,,\\
\dot \eta &= \phantom{-}\omega_c\xi + G(\xi,\eta)\,,
\end{align}
with
\begin{equation}
F(\xi,\eta)=-\varepsilon\xi^3\,,
\qquad
G(\xi,\eta)=\frac{R_h\varepsilon}{\omega_c}\xi^3\,.
\end{equation}

All quadratic derivatives vanish. The only nonzero cubic derivatives relevant to the first Lyapunov coefficient are
\begin{equation}
F_{\xi\xi\xi}=-6\varepsilon\,,
\qquad
G_{\xi\xi\xi}=\frac{6R_h\varepsilon}{\omega_c}\,.
\end{equation}
Using the planar formula, Eq. (\ref{eq:app_explicit_l1}), we obtain
\begin{equation}
l_1
=
-\frac{3}{8}\varepsilon\,.
\end{equation}
Therefore
\begin{equation}
-\frac{\kappa}{l_1}
=
\frac{4}{3\varepsilon}\,.
\end{equation}

\subsection{Leading observable singularity}

Using the general leading-coefficient formula, we obtain
\begin{equation}
a_1
=
\frac{2}{3\varepsilon}\,.
\end{equation}
Therefore
\begin{equation}
\Delta T^2
=
\langle T^2\rangle-T_\ast^2
=
\frac{2}{3\varepsilon}\mu
+
\mathcal O(\mu^2)\,.
\end{equation}

Since \(a_1\neq0\), the ENSO recharge--discharge oscillator belongs to the generic Hopf kink class,
\begin{equation}
n_\ast^{T^2}=1\,.
\end{equation}
This example demonstrates explicitly that the Hopf observable singularity mechanism is not restricted to thermodynamic observables: here the kink appears in the time-averaged climatic variance \(T^2\).


\end{document}